# Local Bifurcations in DC-DC Converters

Chung-Chieh Fang
Taiwan Semiconductor Manufacturing Co., Hsinchu 300, Taiwan
Tel: +886-3-5781688 ext 1261, Fax: +886-3-5637386, Email: ccfangb@tsmc.com.tw
Eyad H. Abed
Dept. of Electrical and Computer Eng. and the Inst. for Systems Research
University of Maryland, College Park, MD 20742 USA, Email: abed@isr.umd.edu

## Keywords
Modelling, DC power supplies

## Abstract
Three local bifurcations in DC-DC converters are reviewed. They are period-doubling bifurcation, saddle-node bifurcation, and Neimark bifurcation. A general sampled-data model is employed to study types of loss of stability of the nominal (periodic) solution and their connection with local bifurcations. More accurate prediction of instability and bifurcation than using the averaging approach is obtained. Examples of bifurcations associated with instabilities in DC-DC converters are given.

## 1 Introduction

There have been many studies of instabilities of DC-DC converters [1, 2, 3, 4, 5]. From a practical perspective, it is useful to classify instabilities depending on how and in what range of operating conditions they arise. Bifurcation theory is a tool that facilitates the study of loss of stability and its implications for dynamical behavior. Upon loss of stability of a steady state solution of a dynamical system, typically a bifurcation occurs in which *new* steady states can arise. Thus, loss of stability of one steady state may lead to operation at a new steady state. A useful classification of bifurcations is that of local bifurcation vs. global bifurcation [6]. In a local bifurcation, the original steady state is an equilibrium point or limit cycle. In a global bifurcation, the original steady state has some other structure (say, an almost periodic solution, or a chaotic orbit). The paper focuses on local bifurcations. It is due to the fact, from a practical point of view, that these bifurcations can be expected to arise before any global bifurcation. Three typical local bifurcations of a periodic orbit are period-doubling (flip) bifurcation, saddle-node bifurcation, and Neimark bifurcation. The associated phenomena are subharmonic oscillation, jump, and quasi-periodicity, respectively.

The most popular model for stability analysis of DC-DC converters has been the averaged model [7]. The averaged model generally makes the following approximations:

1. The nominal steady state is an *equilibrium*.

2. *Equilibrium* stability is of concern.

3. Only *one* steady state is assumed.

4. The duty cycle is a *continuous-time* variable and *unbounded*.

For DC-DC converters, however, the realities are

1. The nominal steady state is a *periodic orbit*, i.e., a limit cycle.

2. *Orbital* stability is of concern. A periodic orbit is stable if any state trajectory starting close to the orbit moves towards the orbit.

3. *Many* periodic and aperiodic steady states may *coexist*.

4. The duty cycle is a *discrete-time* variable and *bounded*.

Close to the onset of instability, the *periodic* nature of the steady state operating condition needs to be considered in order to obtain accurate results. Indeed, it has been reported [8, 9] that averaging leads to erroneous conclusions regarding the onset of instability. Take subharmonic oscillations for example. Period-one and period-two orbits coexist and have the same averaged trajectories. Therefore, the averaged model can not distinguish the two orbits and determine their stability.

This paper employs general sampled-data modeling [10, 11, 12, 13, 14] of DC-DC converters. The bifurcations, indeed exists in real applications, are best explained by the sampled-data model instead of the averaged model. The sampled-data model does not assume those approximations mentioned above. Examples of instability in DC-DC converters are used to illustrate these bifurcations.

The remainder of the paper is organized as follows. In Section 2, local bifurcations of a discrete-time system are summarized. In Section 3, a general model for DC-DC converters developed by the authors in [14, 15] is recalled. The three bifurcations are studied in Sections 4-6. Conclusions are collected in Section 7.

## 2  Local Bifurcations in Discrete-Time Systems

In this section, the basic bifurcation theory used in the paper is recalled. For details, the reader is referred to [16].

Consider a discrete-time parameter-dependent system

$$x_{n+1} = f(x_n, \alpha), \quad x_n \in \mathbf{R}^N, \quad \alpha \in \mathbf{R} \qquad (1)$$

The parameter $\alpha$ is called the bifurcation parameter. Suppose $x = x_0(\alpha)$ is a fixed point of Eq. (1) for all $\alpha$. Denote $A(\alpha) = f_x(x_0(\alpha), \alpha)$, the Jacobian of $f$ with respect to $x$ at $(x_0(\alpha), \alpha)$. The fixed point $x = x_0(\alpha)$ is called a *hyperbolic* fixed point if $A(\alpha)$ has no eigenvalues on the unit circle in the complex plane. If a bifurcation occurs, then it must occur for a value $\alpha_*$ of $\alpha$ for which $A(\alpha)$ is nonhyperbolic. There are three ways in which parameter variation can result in hyperbolicity being violated, and these are associated with three distinct bifurcations:

1. **Period-doubling bifurcation** (the bifurcation associated with a real eigenvalue passing through the value $-1$): There is a curve of fixed point in the $x$-$\alpha$ plane on *both sides* of $\alpha = \alpha_*$ and a curve of period-two points on *one side* of $\alpha = \alpha_*$ intersecting with the first curve at $\alpha = \alpha_*$.

2. **Saddle-node bifurcation** (the bifurcation associated with a real eigenvalue reaching the value 1): There is a unique curve of fixed points in the $x$-$\alpha$ plane passing through $(x_0(\alpha), \alpha_*)$ and locally lying on *one side* of $\alpha = \alpha_*$.

3. **Neimark bifurcation** (the bifurcation associated with a pair of complex conjugate eigenvalues crossing the unit circle): There is a curve of fixed points in the $x$-$\alpha$ plane on *both sides* of $\alpha = \alpha_*$ and the emergence of a small-amplitude "invariant circle" around the fixed-point on *one side* of $\alpha = \alpha_*$.

# 3 Sampled-Data Model for Voltage or Current Mode Control

Without loss of generality, only continuous conduction mode is considered. A summary of the sampled-data modeling of closed-loop DC-DC converters discussed in [14, 15] is given. This model is applicable *both* to voltage mode control and current mode control.

A block diagram model is shown in Fig. 1. In the diagram, $A_1, A_2 \in \mathbf{R}^{N \times N}$, $B_1, B_2 \in \mathbf{R}^{N \times 2}$, $C, E_1, E_2 \in \mathbf{R}^{1 \times N}$, and $D \in \mathbf{R}^{1 \times 2}$ are constant matrices, $x \in \mathbf{R}^N$, $y \in \mathbf{R}$ are the state and the feedback signal, respectively, and $N$ is the state dimension, typically given by the number of energy storage elements in the converter. The source voltage is $v_s$, and the output voltage is $v_o$. The notation $v_r$ denotes the reference signal, which could be a voltage or current reference. The signal $h(t)$ is a $T$-periodic ramp with $h(0) = V_l$ and $h(T^-) = V_h$. In current mode control, it is used to model a slope-compensating ramp. The clock has the same switching frequency $f_s = 1/T$ as the ramp. Within a clock period, the dynamics is switched between the two stages $S_1$ and $S_2$. follows. The system is in $S_1$ immediately following a clock pulse, and switches to $S_2$ at instants when $y(t) = h(t)$.

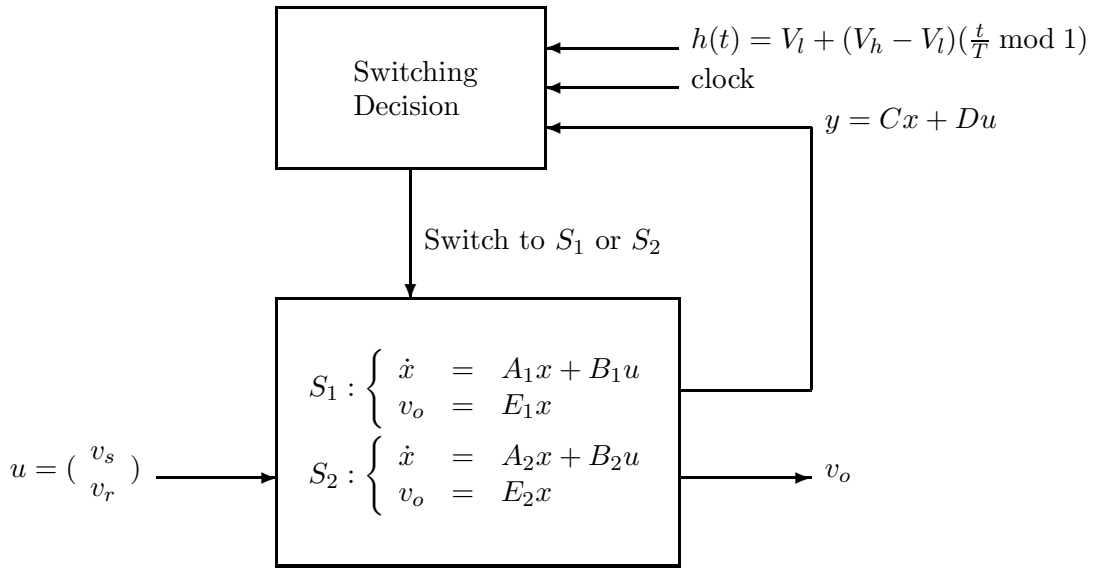

Figure 1: Block diagram model for DC-DC converter operation in continuous conduction mode

Let $x_n = x(nT)$ and steady-state duty cycle $D_c = d/T$. A periodic solution $x^0(t)$ in Fig. 1 corresponds to a fixed point $x^0(0)$ in the sampled-data dynamics. Using a hat ˆ to denote small perturbations (e.g., $\hat{x}_n = x_n - x^0(0)$), The linearized sampled-data dynamics is

$$\hat{x}_{n+1} = \Phi \hat{x}_n \tag{2}$$

where

$$\Phi = e^{A_2(T-d)}(I - \frac{((A_1 - A_2)x^0(d) + (B_1 - B_2)u)C}{C(A_1 x^0(d) + B_1 u) - \dot{h}(d)})e^{A_1 d} \tag{3}$$

Local *orbital* stability of the converter is determined by the eigenvalues of $\Phi$, denoted as $\sigma[\Phi]$. The periodic solution $x^0(t)$ is asymptotically orbitally stable if all of the eigenvalues of $\Phi$ are inside the unit circle of the complex plane.

For comparison, the linearized dynamics of the state-space averaged model is

$$\dot{\hat{x}} = (A_{\text{ave}} + \frac{(A_1 - A_2)X_{\text{ave}} + (B_1 - B_2)uC}{V_h - V_l})\hat{x} \tag{4}$$

where

$$A_{\text{ave}} := A_{\text{ON}}D_c + A_{\text{OFF}}(1-D_c)$$
$$B_{\text{ave}} := B_{\text{ON}}D_c + B_{\text{OFF}}(1-D_c)$$
$$X_{\text{ave}} = -A_{\text{ave}}^{-1}B_{\text{ave}}u$$

# 4 Period-Doubling Bifurcation: Subharmonic Oscillation and Eigenvalue Crossing -1

In the period-doubling bifurcation, a $2T$-periodic solution arises besides the original $T$-periodic solution. In most DC-DC converters, the period-doubling bifurcation is supercritical, where the $2T$-periodic solution is stable and the original $T$-periodic solution becomes unstable. An illustration of such a bifurcation is shown in Fig. 2. Throughout the paper, the stable solution is denoted as solid line; unstable solution, dashed line.

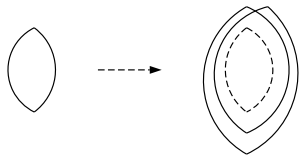

Figure 2: Periodic solution before and after period-doubling bifurcation

Consider the example [2] of a buck converter under voltage mode control shown in Fig. 3. Let $T = 400$ $\mu$s, $L = 20$ mH, $C = 47$ $\mu$F, $R = 22$ $\Omega$, $V_r = 11.3$ V, $g_1 = 8.4$, $V_l = 3.8$, $V_h = 8.2$, (then $h(t) = 3.8 + 4.4[\frac{t}{T} \bmod 1]$), and let $v_s$ be the bifurcation parameter.

Let the state be $x = (i_L, v_C)'$, one has

$$A_1 = A_2 = \begin{bmatrix} 0 & \frac{-1}{L} \\ \frac{1}{C} & \frac{-1}{RC} \end{bmatrix}$$
$$B_1 = \begin{bmatrix} 0 \\ 0 \end{bmatrix} \qquad B_2 = \begin{bmatrix} \frac{1}{L} \\ 0 \end{bmatrix}$$
$$C = \begin{bmatrix} 0 & g_1 \end{bmatrix} \qquad D = \begin{bmatrix} 0 & -g_1 \end{bmatrix}$$
$$E_1 = E_2 = \begin{bmatrix} 0 & 1 \end{bmatrix}$$

The circuit undergoes a series of period-doubling bifurcations beginning at $v_s = 24.5$. The eigenvalues of $\Phi$ (i.e. $\sigma(\Phi)$) as $v_s$ varies from 13.1 to 25.068 is shown in Fig. 4. They are calculated from Eq. (3), while [17] obtains the same graph by numerical estimation. One eigenvalue of $\Phi$ is $-1$ when $v_s = 24.5$, where the period-doubling bifurcation occurs.

After period-doubling bifurcation, the original periodic solution becomes unstable, and a stable $2T$-periodic solution arises. Take $v_s = 26$, for example. The unstable $T$-periodic solution and the stable $2T$-periodic solution are shown in Fig. 5.

The averaged model of this example has been studied in [8] and found to be *stable* for $v_s$ varying from 15 to 40. Therefore the averaged model does not predict the period-doubling bifurcation accurately.

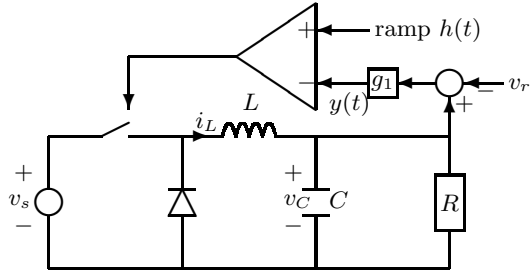

Figure 3: Buck converter under voltage mode control

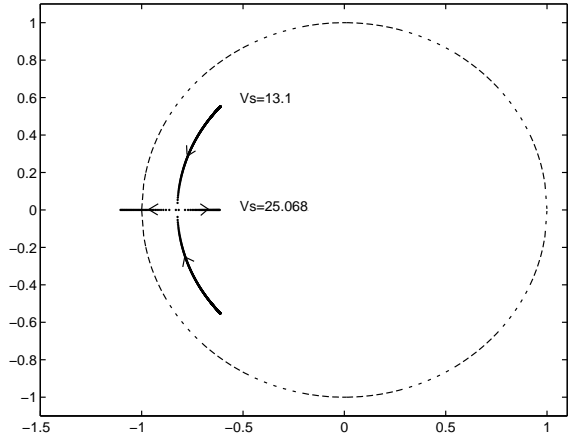

Figure 4: $\sigma(\Phi)$ as $v_s$ varies from 13.1 to 25.068

# 5 Saddle-Node Bifurcation: Jump and Eigenvalue Crossing 1

In the saddle-node bifurcation, a stable $T$-periodic solution collides with an unstable one at the bifurcation point, and no periodic solution exists after the bifurcation. This may explain some jump phenomena, and sudden appearance or disappearance of the nominal periodic solution in DC-DC converters. An illustration of such a bifurcation is shown in Fig. 6.

Consider a buck converter with a discrete-time controller (Fig. 7), where $T = 400$ $\mu$s, $L$=20 mH, $C = 47$ $\mu$F, $R = 22$ $\Omega$. Let $v_s$ be the bifurcation parameter and it is varied from 18.5 V to 20.5 V. For duty cycle 0.7, the nominal inductor current is about $I_p = 0.6785$ and the nominal output voltage is about $V_p = 14.0263$. The switching decision in the cycle, $t \in [nT, (n+1)T)$, is designed as follows (similar to a leading-edge modulation where the switch is off first and then on in a cycle): the switch is turned off at $t = nT$ and turned on at $t = nT + d_n$. The switching instant $d_n$ is updated by $d_n = \ell(0.3T - k_i(i_n - I_p) - k_v(v_n - V_p))$, where $k_i = -8.574 \times 10^{-4}$ and $k_v = 5.53 \times 10^{-5}$ are feedback gains, and $\ell$ is a limiter (to limit the duty cycle within 1):

$$\ell(t) = \begin{cases} 0 & \text{for } t \leq 0 \\ t & \text{for } t \in (0, T] \\ T & \text{for } t > T \end{cases} \qquad (5)$$

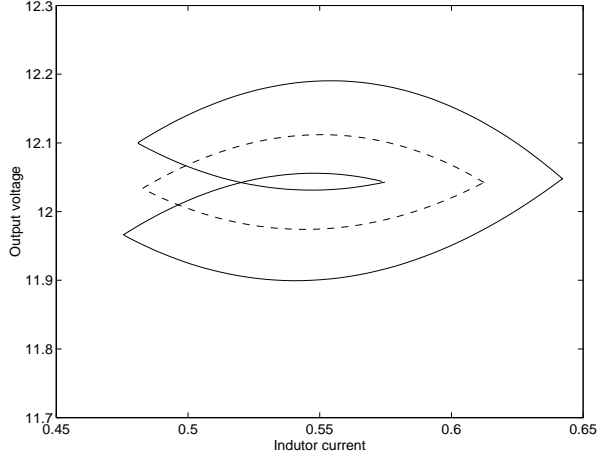

Figure 5: Unstable $T$-periodic solution (dashed line) and stable $2T$-periodic solution for $v_s = 26$

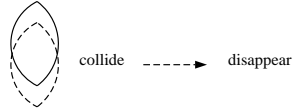

Figure 6: Periodic solution before and after saddle-node bifurcation

The discrete-time control law above leads to new dynamics. It produces different *static* and *periodic* solutions for different $v_s$. First, the switch being always on is a possible operation under some circumstances. When the switch is always on, $d_n = 0$ for any $n$. The steady state solutions are constant instead of being periodic: $v_o(t) = v_s$ and $i(t) = v_s/R$. From Eq. (5), the following inequality needs to hold in order to make $d_n = 0$:

$$\begin{aligned} 0.3T - k_i(i_n - I_p) - k_v(v_n - V_p) &= 0.3T - k_i(\frac{v_s}{R} - I_p) - k_v(v_s - V_p) \\ &= 0.3T + k_i I_p + k_v V_p - (\frac{k_i}{R} + k_v)v_s \\ &\leq 0 \end{aligned}$$

Therefore for $v_s > (\frac{k_i}{R} + k_v)/(0.3T + k_i I_p + k_v V_p) = 19.213$, the switch can be always on.

However, the switch being always on is not the *only* possible operation for $v_s > 19.213$. For $v_s \in (19.213, 20)$, there are another two periodic solutions: one is stable, another one is unstable.

Take $v_s = 19.9$, for example. One *stable* periodic solution with duty cycle 0.6267 and one *unstable* periodic solution with duty cycle 0.7878 are shown as the solid line and dashed line respectively in Fig. 8. The stable one has output voltage around 12.5 V; unstable one has output voltage around 15.7 V. As $v_s$ is further increased, these two periodic solutions move closer and collide when $v_s = 20$. For $v_s = 20$, one eigenvalue of $\Phi$ is 1 and a saddle-node bifurcation occurs.

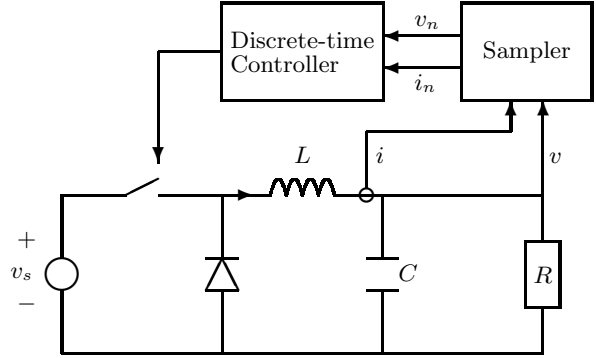

Figure 7: Buck converter under discrete-time control

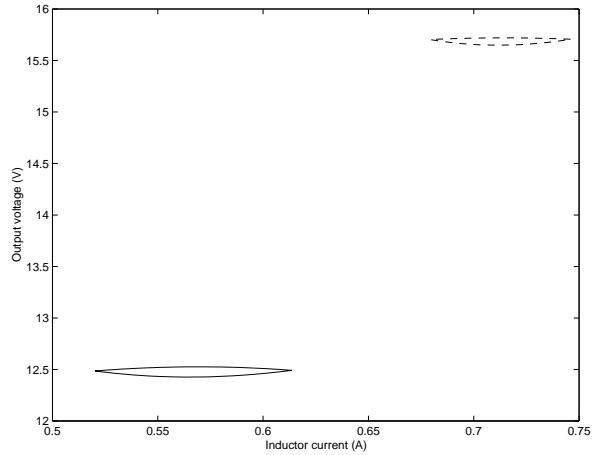

Figure 8: Stable periodic orbit (solid line) and unstable periodic orbit (dashed line) for $v_s$=19.9

If $v_s$ is increased a little bit above 20, the operation suddenly *jumps* to the situation where the switch is always on and the output voltage jumps from 14 V to 20 V.

The circuit is simulated for $v_s \in [18.5, 20.5]$ and the resulting bifurcation diagram is shown Fig. 9. In the figure, the upper solid line is for the stable *static* solution when the switch is always on (hence duty cycle is 1 and $d_n = 0$); the dashed line and the lower solid line are for unstable and stable *periodic* solutions respectively with duty cycle less than 1. For $v_s$ below 19.213, there is only one stable periodic solution and the output voltage is regulated below 11 V.

# 6 Neimark Bifurcation: Quasi-Periodicity and Eigenvalues Crossing Unit Circle

When a Neimark bifurcation occurs, the original periodic solution (with frequency $f_s$) is modulated by another frequency, $\frac{f_s}{2\pi}\angle\sigma(\Phi)$, i.e., $\frac{f_s}{2\pi}$ times the argument (i.e., phase) of the pair of eigenvalue of $\Phi$ crossing the unit circle. The state trajectory will be periodic (phase-locking) if these two frequencies are commensurate; otherwise it will be quasiperiodic.

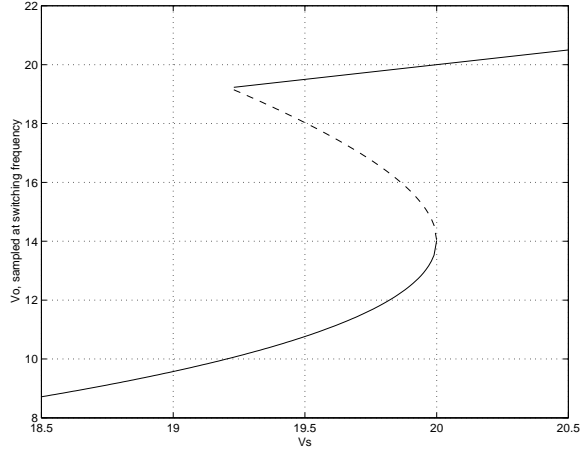

Figure 9: Bifurcation diagram showing saddle-node bifurcation at $v_s = 20$

Consider the example [18, p.228] of a buck converter under voltage mode control shown in Fig. 10. The system parameters are $f_s=$ 15 kHz, $L$=0.9 mH, $C$=22 $\mu$F, $R$=20 $\Omega$, $v_r$=5 V, $R_1 = R_2$=7.5 k$\Omega$, $R_3$=60 k$\Omega$, $C_2$=0.4 $\mu$F, $v_s$=30 V, $V_L$=2.8 V, $V_U$=8.2 V, (then $h(t) = 2.8 + 5.4[\frac{t}{T} \bmod 1]$). All parasitic resistances are ignored. Here under the voltage mode control, the output voltage is sensed, scaled by 0.5, compared with $v_r = 5$, and input into the error amplifier. Therefore, an output voltage around 10 V is expected.

Let the state $x = (i_L, v_C, v_{c2})'$, one has

$$A_1 = A_2 = \begin{bmatrix} 0 & \frac{-1}{L} & 0 \\ \frac{1}{C} & \frac{-1}{RC} & 0 \\ 0 & \frac{1}{R_1 C_2} & \frac{-1}{R_3 C_2} \end{bmatrix}$$

$$B_1 = \begin{bmatrix} \frac{1}{L} & 0 \\ 0 & 0 \\ 0 & \frac{-1}{C_2}(\frac{1}{R_1} + \frac{1}{R_2}) \end{bmatrix} \quad B_2 = \begin{bmatrix} 0 & 0 \\ 0 & 0 \\ 0 & \frac{-1}{C_2}(\frac{1}{R_1} + \frac{1}{R_2}) \end{bmatrix}$$

$$C = \begin{bmatrix} 0 & 0 & -1 \end{bmatrix} \quad D = \begin{bmatrix} 0 & 1 \end{bmatrix}$$

$$E_1 = E_2 = \begin{bmatrix} 0 & 1 & 0 \end{bmatrix}$$

The fixed point $x^0(0)$ is $(0.25, 10, 0.39)'$. The eigenvalues of $\Phi$ are 0.8799 and $0.8797 \pm 0.4474i$, which are inside the unit circle. Thus the periodic solution is asymptotically orbitally stable.

As $v_s$ is increased from 30 V, the magnitude of the complex pair of eigenvalues begins to grow. For $v_s$=36.9, the eigenvalues ($0.8897 \pm 0.4567i$) exit the unit circle. Thus a Neimark bifurcation occurs. A low oscillating frequency $\frac{f_s}{2\pi} \angle (0.8897 + 0.4567i)$= 1132 Hz modulating the original one $f_s$ is expected ($\angle$ denotes the angle in radian). Since these two frequencies are not commensurate, the steady state is quasiperiodic.

For $v_s$=30 (before bifurcation), the stable periodic solution $x^0(t)$ is shown as a solid line in Fig. 11. For $v_s$=50 (after bifurcation), the periodic solution becomes unstable (dashed line in Fig. 11). A quasiperiodic state trajectory arises (Fig. 12), coexisting with the unstable periodic solution.

Output voltage waveforms of the quasiperiodic steady state and the unstable periodic solution are shown as solid line (with larger amplitude) and dashed line respectively in Fig. 13 for $v_s$=50. From the figure, the quasiperiodic steady state has two oscillating frequencies as expected: $f_s$ modulated by a lower frequency around 1132 Hz.

Next, the state-space averaged model (4) is analyzed. For $v_s = 36.9$, a pair of eigenvalues, $0.3 \pm 7113.5i$, becomes unstable. The modulating frequency is $7113.5/2\pi = 1132$ Hz, which also gives correct prediction. This is expected because the $T$-periodic solution can be averaged to an equilibrium and the quasiperiodic solution can be averaged to a low frequency (1132 Hz) oscillation.

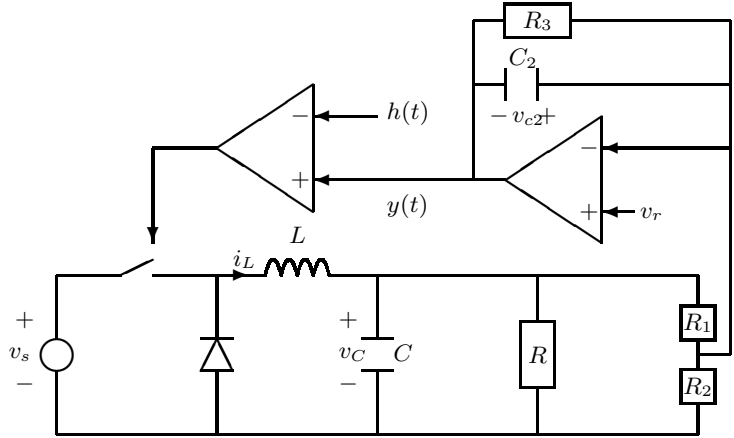

Figure 10: Buck converter under voltage mode control

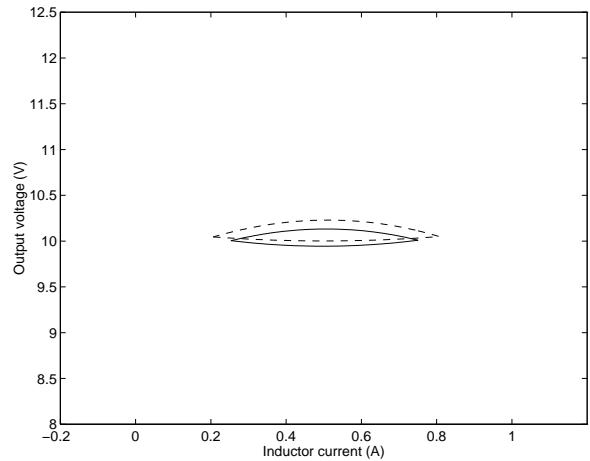

Figure 11: Stable periodic orbit (solid line) for $v_s$=30 becomes unstable (dashed line) for $v_s$=50

# 7 Conclusion

Local bifurcations in DC-DC converters are studied using sampled-data models. The bifurcations considered are period-doubling bifurcation, saddle-node bifurcation, and Neimark bifurcation. Physical phenomena associated with these bifurcations are subharmonic oscillation, jump, and quasi-periodicity, respectively, and they are explained in details. Orbital stability is emphasized.

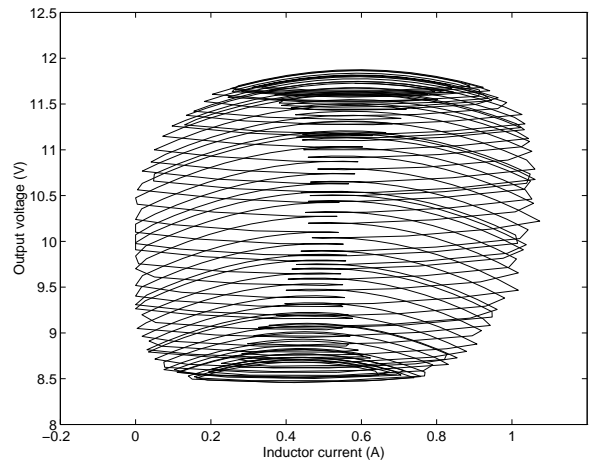

Figure 12: Quasiperiodic state trajectory in state space for $v_s$=50

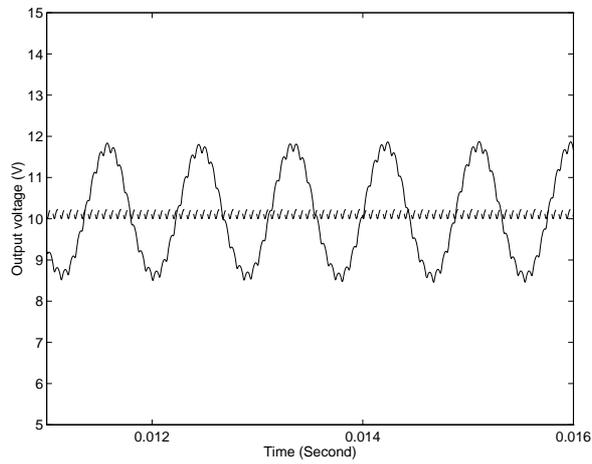

Figure 13: Quasiperiodic output (larger amplitude) and unstable $T$-periodic output for $v_s$=50

Instabilities in DC-DC converters can be related to these bifurcations. These bifurcations, indeed exists in real applications, are well explained by the sampled-data model. The only bifurcation found predicted by the state-space averaged model is the Neimark bifurcation.